\documentclass[10pt,a4paper]{article}
\usepackage[left=14mm, top=10mm, right=14mm, bottom=20mm, nohead, foot=10mm]{geometry}
\usepackage[affil-it]{authblk}

\let\OLDthebibliography\thebibliography
\renewcommand\thebibliography[1]{
  \OLDthebibliography{#1}
  \setlength{\parskip}{0pt}
  \setlength{\itemsep}{0pt plus 0.3ex}
}

\usepackage{multicol}

\usepackage{caption}

\usepackage{misccorr}

\usepackage[english]{babel}
\usepackage{amsmath,amssymb}
\usepackage{textcomp}

\usepackage{xcolor}

\usepackage{hyperref}
\hypersetup{colorlinks=true, linkcolor=[rgb]{0.6 0 0}, citecolor=[rgb]{0 0.4 0}, urlcolor=[rgb]{0 0 0.6}}

\usepackage{graphicx}
\usepackage{epsfig}
\newenvironment{Figure}
  {\par\medskip\noindent\minipage{\linewidth}\captionsetup{type=figure}}
  {\endminipage\par\medskip}
\usepackage{epstopdf}

\title{\Large \textbf{New type of stable particle like states in chiral magnets \\ (\textit{Chiral bobbers})}}
\author[1]{\large F.\,N. Rybakov\thanks{f.n.rybakov@gmail.com}}
\author[1]{\large A.\,B. Borisov}
\author[2]{\large S. Bl\"ugel}
\author[2]{\large N.\,S. Kiselev\thanks{n.kiselev@fz-juelich.de}}
\affil[1]{\normalsize M.\,N. Miheev Institute of Metal Physics of Ural Branch of Russian Academy of Sciences, Ekaterinburg 620990, Russia}
\affil[2]{\normalsize Peter Gr\"unberg Institut and  Institute for Advanced Simulation, Forschungszentrum J\"ulich and JARA, D-52425  J\"ulich, Germany}

\date{}

\begin{document}

%{\let\newpage\relax\maketitle}

\maketitle
\vspace{-3.0\baselineskip}

\renewcommand{\abstractname}{\vspace{-\baselineskip}}
\begin{abstract}
We present a new type of a thermodynamically stable magnetic state at interfaces and surfaces of  chiral  magnets. 
The state is a  soliton solution of  micromagnetic equations localized in all three dimensions near a boundary and contains a singularity, but nevertheless has a finite energy.
Both features combine to a quasi-particle state for which we expect unusual  transport and dynamical properties.
It exhibits high thermal stability and thereby can be considered as promising object for fundamental research and practical applications in spintronic devices.  
We provide arguments that such a state can be found in different B20-type alloys \textit{e.g.}\  Mn$_{1-x}$Fe$_x$Ge,  Mn$_{1-x}$Fe$_x$Si, Fe$_{1-x}$Co$_x$Si.  
\end{abstract}

%\thispagestyle{empty}

%\section*{Introduction}
%{\let\newpage\relax\twocolumn[]}

\begin{multicols}{2}[]
Many branches of modern physics focus on the systems, which exhibit a rich variety of metastable states.
Special attention is drawn to systems characterized by high energy barrier, which are able to persist in a particular metastable state for an unbounded or noticeably long time.
Such systems are within the scope of interest for optical, magnetic and electronic data storage technology. 
Basically, almost all modern electronics and spintronics are based on handling and manipulating such metastable states.
For instance, the magnetic domains in magnetic data storage devices are the best examples of such metastable states separated by high-energy barriers.
In hard disk drives high energy barriers are provided by a high magneto-crystalline anisotropy of the material, which makes at the same time the magnetic domains moveless and affixed to their positions.
The disadvantage of such data storages is that the detector has to be moved to the place where the data bit is stored or the whole storage media has to be moved towards the detector.

Modern spintronics offers  many alternative ideas for the next-generation magnetic data storage devices that are based on utilizing movable metastable states.
The most promising candidates are domain walls in nanowires \cite{Parkin_15} and localised magnetic vortices \cite{Bogdanov_89} also known as chiral magnetic skyrmions \cite{Kiselev_11,Fert_13}.
The latter appear in magnets with Dzyaloshinskii-Moriya interaction (DMI) \cite{Dzyaloshinskii,Moriya} as an isolated metastable state or as a ground state, when skyrmions condense into a lattice. 

From the point of view of the field theory, skyrmions are two-dimensional solitons known as an exception to the Hobart-Derrick theorem \cite{Rajaraman} due to the  presence of Lifshitz invariants \cite{Bogdanov_95}.
Due to their unusual transport properties,
and compact size, which at some conditions can be reduced up to a few nano-meters \cite{Shibata_13} or even a few atomic distances \cite{Romming_13},
such types of chiral votices are considered as promising particle for information technology as well as an interesting object for fundamental research in non-linear physics. 

In this paper, we present a new type of metastable particle-like state -- a hybrid particle composed of a smooth magnetization vector field and a magnetic singularity.
It is a three-dimensionally localized soliton of the nonlinear equations for a unit vector field.
Such a quasi-particle is more compact and it turns out to be energetically more favorable than the chiral magnetic skyrmion in a wide range of parameters.

%\section*{Model}-----------------------------
Our approach is based on a classical spin model of the simple cubic lattice described by the following Hamiltonian \cite{Yu_10} consisting of three contributions: the Heisenberg exchange interaction, the Dzyaloshinskii-Moriya interaction and Zeeman term 
\begin{equation}
E\!=\!- J \sum_{\left\langle ij\right\rangle }
 \mathbf{n}_i  \cdot  \mathbf{n}_j - 
\sum_{\left\langle ij\right\rangle } 
\!\mathbf{D}_{ij}  \cdot  [\mathbf{n}_i\! \times\!  \mathbf{n}_j]   - \! \mu_\mathrm{s} \mathbf{H}\!\sum_{i}\mathbf{n}_i,         
\label{E_tot}
\end{equation}
where $\left\langle ij\right\rangle$ denote summation over all nearest-neighbour pairs, $\mathbf{n}_i=\mathbf{M}_i/\mu_\textrm{s}$ is a unit vector of the magnetic moment at lattice site $i$,
$J$ is the exchange coupling constant and $\mathbf{D}_{ij}$ is the Dzyaloshinskii-Moriya vector defined as $\mathbf{D}_{ij}=D \, \mathbf{r}_{ij}$, 
$D$ is the DMI scalar constant and $\mathbf{r}_{ij}$ is a unit vector pointing from site $i$ to site $j$,
$\mathbf{H}$ is an external magnetic field applied along $z$-axis, perpendicular to the film plane.
Note, we assume isotropic DMI which means that the value $|\mathbf{D}_{ij}|$ is fixed for all three spatial directions.
Periodical boundary condition is applied in $xy$-plane.

In order to preserve the generality of the results of our discrete model and to be able to compare them to the earlier developed theory based on continuum approximation \cite{Bogdanov_89, Bogdanov_94, Bogdanov_99, Bogdanov_06, Bogdanov_11} we use the following notations
\begin{eqnarray}
L_\mathrm{D}=4\pi\frac{\mathcal{A}}{\mathcal{D}}=2\pi a\frac{J}{D},\ \
H_\mathrm{D}=\frac{\mathcal{D}^2}{2M_\mathrm{s}\mathcal{A}}=\frac{D^2}{\mu_\mathrm{s}J} ,
\label{LH_params}
\end{eqnarray}
where $L_D$ is the lowest period of incommensurate spin spiral which exists below the critical field $H_\mathrm{D}$; $\mathcal{A}$ and $\mathcal{D}$ are micromagnetic constants of exchange and DMI respectively;
$M_\mathrm{s}$ is the magnetization of the material; $a$ is the lattice constant. 

We use a nonlinear conjugate gradient method with adaptive stereographic projections implemented on NVIDIA CUDA architecture for direct minimization of the Hamiltonian (\ref{E_tot})\cite{Supplement}.

\begin{Figure}
\includegraphics[width=1.0\columnwidth]{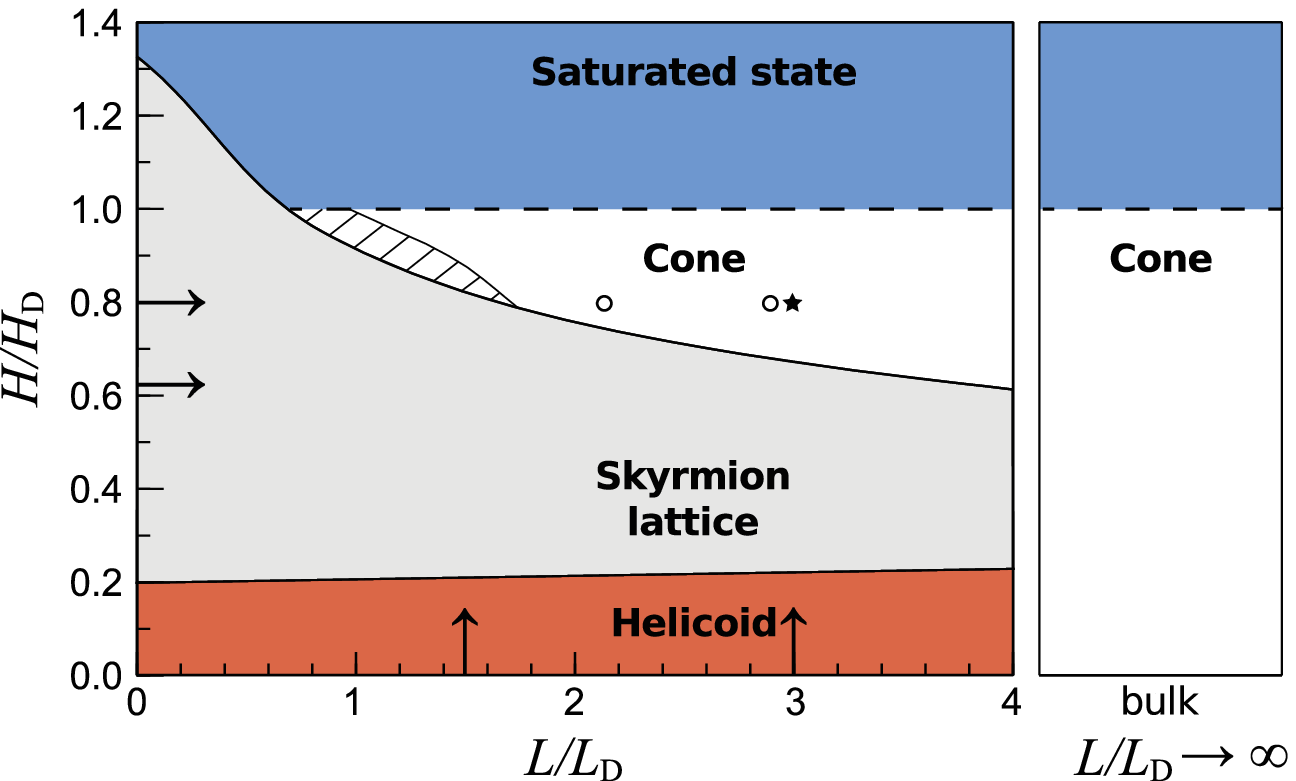}
\caption{\small Phase diagram of the ground state for isotropic helimagnetic thin film of the thickness $L$ at magnetic field $H$ applied normally. 
The dashed area within the conical ground state corresponds to the isolated skyrmion as the lowest energy metastable state.
Arrows indicate the values of the fixed parameters used in the energy calculation presented in Figs.~\ref{energies}a-d. 
The star symbol corresponds to the
parameters used for the calculation of non-zero anisotropy
case, see Figs.~\ref{energies}e, f; the circles - for the calculation of the energy barriers, see Fig.~\ref{MEP}.
}
\label{PD}
\end{Figure}

%\section*{Results}-----------------------------
In order to identify the range of parameters corresponding to the stable configurations we have calculated the magnetic phase diagram for the
ground state, see Fig.~\ref{PD}.
Here, solid lines correspond to the first order phase transition between helical state \cite{helix}, skyrmion lattice (SkL), conical phase, and saturated state. 
Horizontal dashed line between conical and saturated states corresponds to the second order phase transition. 
Under the condition $L_\mathrm{D}$ is one order of magnitude or more larger than $a$, discrete (\ref{E_tot}) and continuum \cite{Supplement} approaches are consistent.

Results presented in Fig.~\ref{PD} are consistent with the continuum theory  \cite{Rybakov_13} and experimental data on the direct observation of SkL in thin films of cubic helimagnetic alloys as 
Fe$_{1-x}$Co$_x$Si \cite{Yu_10}, Mn$_{1-x}$Fe$_x$Ge \cite{Shibata_13},  Mn$_{1-x}$Fe$_x$Si \cite{Yokouchi_14} and pure MnSi \cite{Yu_15} and FeGe \cite{Yu_11}.
The mechanism of SkL stabilization in thin films of cubic helimagnet differs from two-dimensional systems with interface induced DMI \cite{Dupe_14} and anisotropic bulk helimagnets \cite{Bogdanov_14,Leonov_PhD}.
According to Ref.~\cite{Rybakov_13}, the free boundary of the layer which provides an additional degree of freedom to the magnetic system and changes the common energy balance are responsible for the SkL stabilization.
In particular, due to the lack of exchange coupled neighbors on the surface the spin structure of an isolated skyrmion tube (SkT) exhibits a small twist of magnetization with respect to the surface normal, as schematically shown in Fig.~\ref{scheme}a.
Such an induced twist propagates through the whole thickness of the sample, accumulates the energy gain of DMI contribution along the $z$-axis and results in significant reduction of the total energy of the SkT.
As a result, within a certain range of applied field the energy of the SkL becomes lower than the energy of the conical phase.
The equilibrium value of the angle for such a surface twist is defined by the competition between positive contributions of exchange coupling and two negative contributions: i) in-plane DMI - between the spins in the same plane and ii) out-of-plane DMI - between the spins along $z$-axis.
For the SkT and ChB this angle $\varphi_0\lesssim 15^\circ$ (see Fig.\ref{scheme}).
Recently the effect of the chiral surface twist has been proven experimentally \cite{Surf_twist1,Surf_twist2}.

\begin{Figure}
\includegraphics[width=1.0\columnwidth]{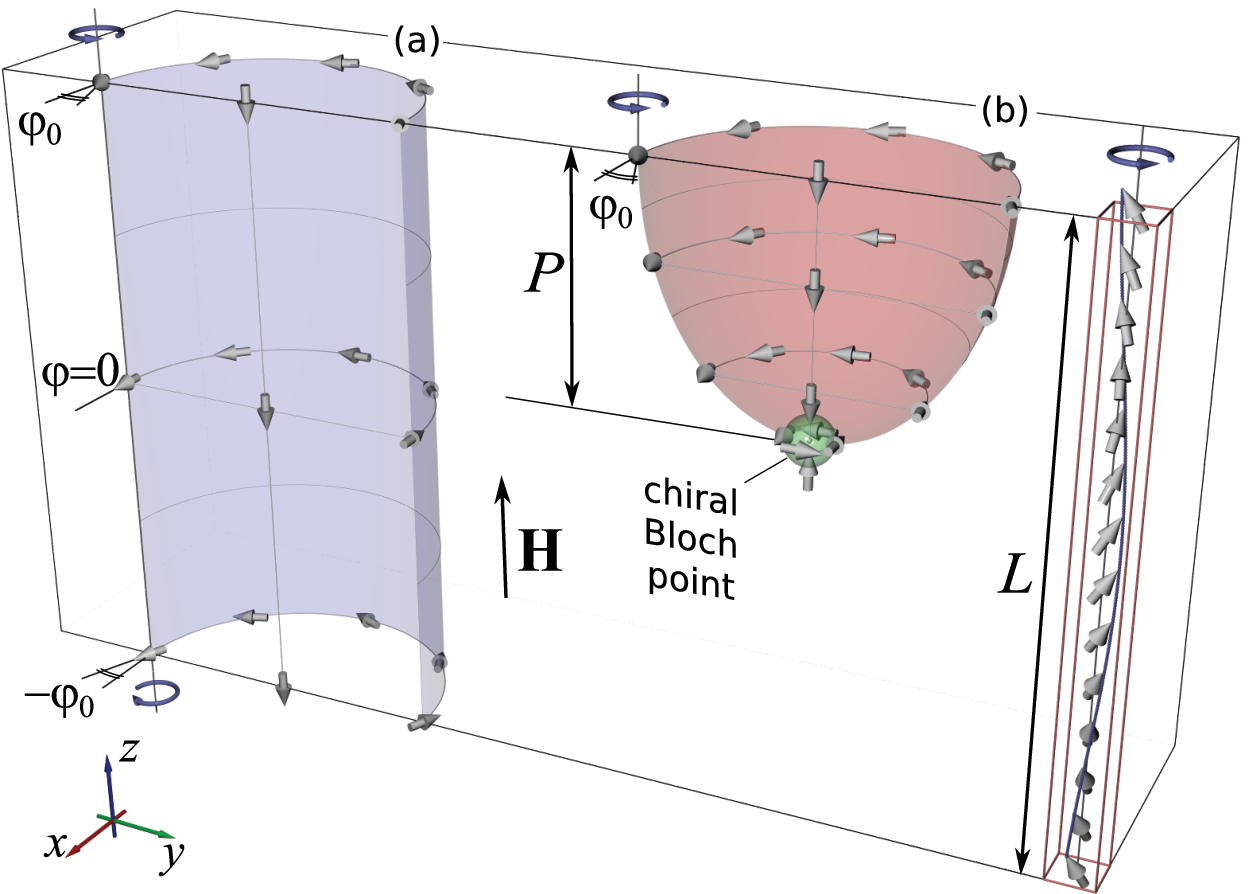}
\caption{\small Schematic representation of the vector field and cross sections of corresponding isosurfaces ($n_\mathrm{z}=0$) for skyrmion tube (SkT) (a) and chiral bobber (ChB) (b). The small (green) sphere indicates the position of the chiral Bloch point. The spins in red marked box on the right illustrate the magnetization distribution in the conical phase, which surrounds these localized metastable states. Circular blue arrows indicate the sense of magnetization rotation within the conical phase and surface induced twist.  
For the exact spin structure of SkT and ChB see Supplementary Movies~1 and 2~\cite{Supplement}.
}
 \label{scheme}
\end{Figure}

Everywhere within the conical phase the lowest energy metastable state corresponds to the that shown in Fig.~\ref{scheme}b.
The exception is the small dashed area in Fig.~\ref{PD} where SkT dominates.

The vector field corresponding to such a solution is characterised by the shape of the isosurface $n_z=0$, which is close to a shape of paraboloid, see Fig.~\ref{scheme}.
An essential element of the corresponding magnetic structure is a singularity situated at a finite distance $P$ from the surface, see the green sphere at the bottom of the red isosurface.
This type of singularity is known in ferromagnets as Bloch point (BP) or hedgehog \cite{Malozemoff}, and in other condensed matter systems e.g. ultra cold gas \cite{Ray_14} and spin ice \cite{Castelnovo_08} as magnetic monopole.
In helimagnets, due to the presence of DMI, the vector field around the singularity exhibits a chiral structure. 
Thus we use the name chiral BP to refer to it.
Earlier it has been shown that some dynamical processes in chiral magnets are accompanied by the appearance of this kind of singularities \cite{Milde, Rosch}.
We succeeded to get an analytical expression for the vector field in the close vicinity to the chiral BP, which gives an opportunity to estimate its energy contribution to the total energy of the system \cite{Supplement}.

In the cross section with the plane parallel to the layer surface its spin structure mimics the magnetization distribution of skyrmion with the diameter diminishing with the distance from the surface.
We found that the value of the penetration depth always satisfy the condition $P \le L_\mathrm{D}/2$ with an accuracy up to $\pm a/2$.
Because of the essential chirality of such a state and its localization close to the surface with the finite penetration depth, like a fishing bobber at a water surface, we use term \textit{chiral bobber} (ChB) to refer to this object.
\end{multicols} 

\begin{Figure}
\includegraphics[width=1.0\textwidth]{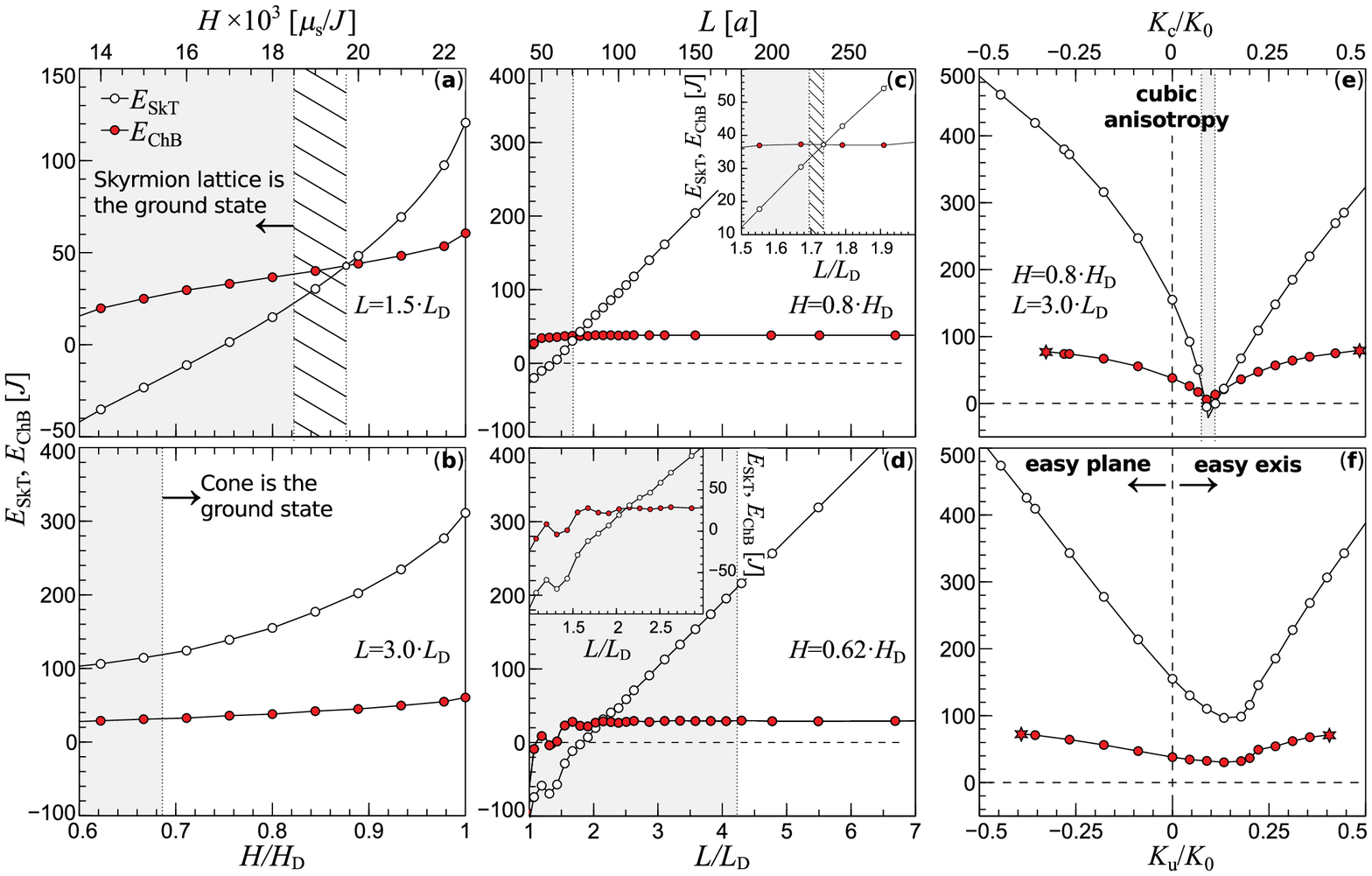}
\caption{\small Energy dependencies for isolated skyrmion tube (SkT, open circles) and chiral bobber (ChB, red solid circles) on applied magnetic field for fixed film thickness (a)-(b), on film thickness for fixed applied field (c)-(d), and as a function of varying values of cubic (e) and uniaxial anisotropy (f) for fixed thickness and field. The energies are presented relative to the pure conical phase.
Note, the cases in (a)-(d) correspond to the isotropic case, $K_\mathrm{c}=K_\mathrm{u}=0$. 
The top and bottom scales in (a-b) and (c-d) are identical but given in absolute and reduced units, respectively.
For these calculations we used $J=1$, $D=0.15$ ($L_\mathrm{D}=41.89a$) and a size of the simulated domain of $256\times 256\times L_n$ spins, where $L_n=40\div280$.}
\label{energies}
\end{Figure}

\begin{multicols}{2}[] 
We calculated the energy dependencies of the ChB and SkT at applied field and varying thickness, see Figs.~\ref{energies}a-d. 
Here, we use the same color notation as in the phase diagram, Fig.~\ref{PD}: white and grey areas correspond to the stability range of conical phase and SkL, respectively, dashed area denotes the range within the conical ground state where the SkT is lower in energy than the ChB. 
As follows from the Figs.~\ref{energies}a-d almost everywhere within the conical phase the ChB has energy much lower than the SkT.
The energy difference between the ChB and SkT becomes most pronounced by  varying the thicknesses, see Figs.~\ref{energies}c, d.
The energy of the SkT increases linearly with the thickness.
At the same time, due to the localisation of the ChB near the surface its energy is almost thickness independent.
It shows significant variation only at low fields and very small thickness, outside the range of the conical phase stability, see insets in Fig.~\ref{energies}d.

In order to estimate the stability of the ChB in anisotropic systems we add to the model Hamiltonian Eq.(\ref{E_tot}) the following term
\begin{eqnarray}
E_\mathrm{K}\!=\!-\!\sum_{i}
\left\lbrace 
 K_\mathrm{u} {n}^2_{i,{z}}
 +  K_\mathrm{c} \left[ n^4_{i,{x}} + n^4_{i,{y}} + n^4_{i,{z}}\right]
\right\rbrace ,
\end{eqnarray}
where $n_{i,\mathrm{x}}$ is $x$-component of unity vector $\mathbf{n}$ at the lattice site $i$, $K_\mathrm{c}$ is the cubic anisotropy constant and $K_\mathrm{u}$ is the constant of uniaxial anisotropy induced in the system, e.g. by the interfaces or under external stress.
In Figs.~\ref{energies}e, f the energies of the ChB and SkT are presented as functions of varying anisotropy constants at fixed applied field $H=0.8H_\mathrm{D}$ and thickness $L=3L_\mathrm{D}$.
We use reduced units of anisotropy where $K_{0}=a^3\mathcal{D}^2/(4\mathcal{A})=D^2/(2J)$.

As follows from the figures, in the whole range of existence within the conical phase the ChB remains energetically more favorable than the SkT.
The star symbols in Figs.~\ref{energies}e and f corresponds to the collapse of the ChB.
In Fig.~\ref{energies}e the narrow gray region denotes the range of SkL stability, which is consistent with previous theoretical study \cite{Leonov_PhD}. 

Due to the presence of free boundaries and singularities as an essential element of the solution the topology of such object should be described in frame of relative homotopy groups \cite{Vol1978}.
Since, our boundary conditions impose no restrictions on the magnetization, as an order parameter, the corresponding group is trivial, $\pi_2(\mathbb{S}^2,\mathbb{S}^2)=0$.
It means that the singularity can be pushed out on the surface and then smoothed. 
It is important to emphasize that when certain restrictions to the order parameter are imposed at the boundary it may lead to formation of topologically nontrivial states, as in case of a boojum in A-phase in $^3$He \cite{Monastyrsky,Volovik}.
Despite the fact that solution for the ChB is non-topological, it shows the high energy barrier.  

In order to estimate the energy barriers we have calculated the minimum energy
path (MEP) with the geodesic nudged elastic band (GNEB) method \cite{Bessarab_15}, which is an extension of the original nudged elastic band method~\cite{jonsson_98,Henkelman_00} to magnetic systems.
In the GNEB method, the curvature of the configurations space of a magnetic system arising from a constraint on the length of magnetic moments is taken into account.
The path between the metastable states is presented by the discrete sequence of system snapshots called images.
These images are then brought to the MEP in the curved configuration space.

\begin{Figure}
\includegraphics[width=1.0\textwidth]{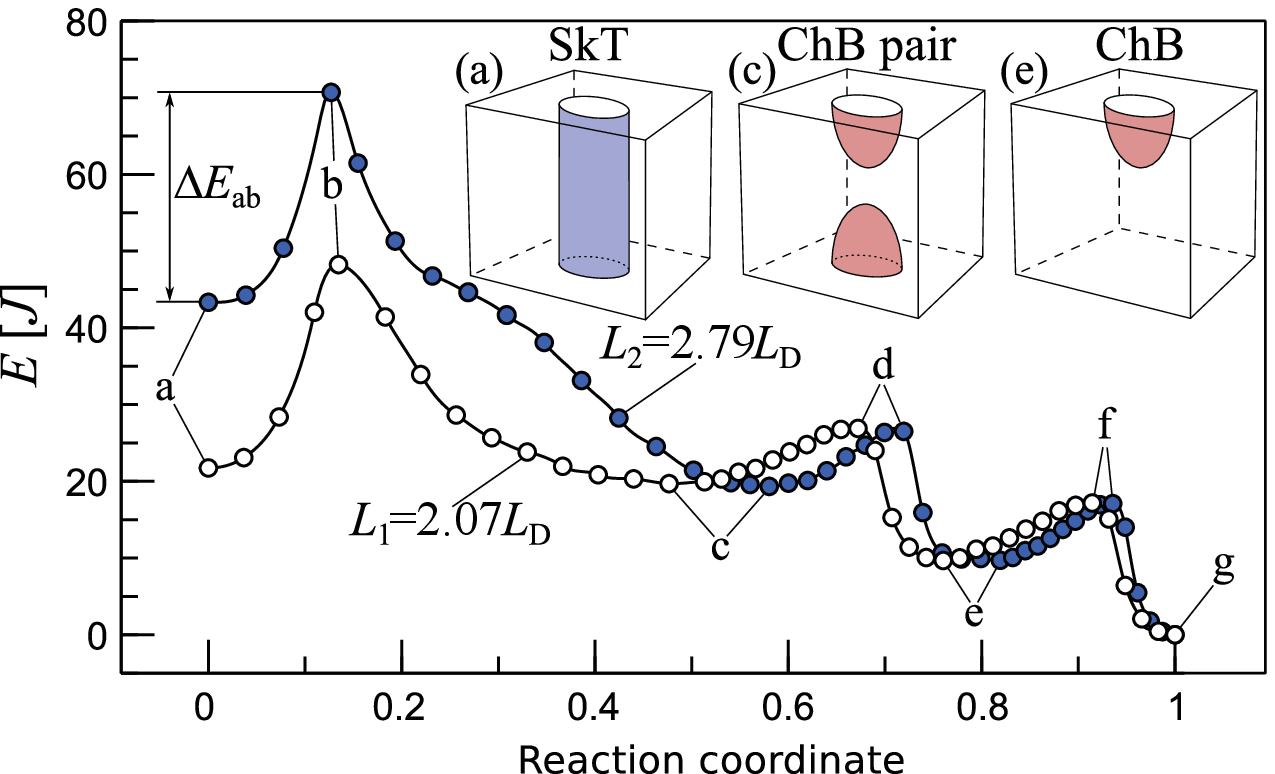}
\caption{\small Minimum energy path calculated for $L_1=2.07L_\mathrm{D}$, and $L_2=2.79L_\mathrm{D}$ at $H/H_\mathrm{D}=0.8$, $K_\mathrm{c}=K_\mathrm{u}=0$.
Reaction coordinate is an order parameter which has a meaning of the relative distance between the states in the configuration space \cite{Bessarab_15}.  
The local minima $a$, $c$, and $e$ correspond to the energies of SkT, pair of ChBs, and single ChB, respectively, see also schematic representation in the corresponding insets. 
The difference between the energy values in the local minimum and nearest maximum corresponding to the saddle points (see $b$, $d$, and $f$) defines the energy barrier responsible for the stability of corresponding state. The reference level ($E=0$) is the energy of the global minimum $g$ corresponding to the pure conical phase \cite{Supplement}.}
 \label{MEP}
\end{Figure}

Examples of MEPs are presented in Fig.~\ref{MEP}.  
For this calculations we used the following parameters: $J=1$, $D=0.45$ ($L_\mathrm{D}=13.96 a$) and domain size of $30\times30\times30$  and $30\times30\times40$.
The elimination of the SkT occurs via nucleation of a pair of chiral BPs (point $b$) moving towards opposite surfaces and results in appearance of two repulsive ChBs (point $c$) which have energy higher than two isolated ChB.

As follows from Fig.~\ref{MEP}, the energy barriers for the ChB and the SkT have comparable height, $\Delta E_\mathrm{ef}=0.43\Delta E_\mathrm{ab}$ for $L_\mathrm{1}$ and  
$\Delta E_\mathrm{ef}=0.28\Delta E_\mathrm{ab}$ for $L_\mathrm{2}$.
From that we may conclude that the thermal stability range for ChB is expected to be of the same order of magnitude as for isolated SkT.
Moreover, due to the high activation energy required for the SkT nucleation and complexity of the MEP, we may expect that the probability of spontaneous nucleation of the skyrmions driven by thermal fluctuation at least in this regime has to be much lower in comparison to the ChB.
Indeed, $\Delta E_\mathrm{gb}=2.8\Delta E_\mathrm{gf}$ for $L_{1}$, and $\Delta E_\mathrm{gb}=4.1\Delta E_\mathrm{gf}$ for $L_{2}$.
This simple argument is confirmed by the results of our Monte Carlo simulation.
In particular, for the parameters $L=2.36L_\mathrm{D}$ and $H=0.8H_\mathrm{D}$ the spontaneous nucleation of the ChBs has been observed during the simulated annealing, while the spontaneous nucleation of the SkT has not been observed in this regime at all \cite{Supplement}.

%\section*{Conclusions}-----------------------------
In conclusion, we have shown that for isotropic and weakly anisotropic helimagnets  the chiral bobber corresponds to the lowest energy metastable state in a wide range of thicknesses and applied fields.
It exhibits high thermal stability, which supports our prediction to observe such a state at the surface of bulk chiral magnets, in thin films, or heterostructures of chiral magnets for thicknesses $L\geq 0.6 L_\mathrm{D}$.
The most promising candidate material offering a  direct observation of chiral bobbers seem to be Mn$_{1-x}$Fe$_x$Ge, where the period of the spin spiral $L_\mathrm{D}$ can be controlled in a wide range ($3$-$160$ nm) by varying concentration $x$ of Fe \cite{Shibata_13}.
Because of the three-dimensional nature of the object, off-axis electron holography in a transmission electron microscope is a promising technique visualizing its three-dimensional magnetic structure \cite{Park_14,Midgley}. 

The chiral bobber constitutes a new class of particles -- the hybrid particles composed of a smooth magnetization field and a magnetic singularity. Both contribute to three-dimensional torque fields and interesting dynamical and electron transport properties. 

We speculate that the chiral Bloch point contributes to a non-zero scalar chirality that produces an orbital moment which can couple to optics. 
This new type of particle may offer the opportunity to  introducing atomic-scale singularities as a new concept in spintronic device design.

\vspace{0.5\baselineskip}

Authors thank P.\,F. Bessarab for his help with the implementation of the GNEB method and fruitful discussion of the paper. 
The work of F.\,N.\,R. was supported
by RFBR, Research Project No. 14-02-31012.
\end{multicols} 
\vspace{0.5\baselineskip} 
 
\begin{multicols}{2}[] 
\renewcommand{\section}[2]{}

\end{multicols}

\setcounter{equation}{0}
\setcounter{figure}{0}

\renewcommand{\thesection}{\text{A}\arabic{section}}
\numberwithin{equation}{section}
\numberwithin{figure}{section}

\part*{\Large Appendices}
\label{app}

\section{Micromagnetic analysis for point singularities in isotropic chiral magnets}
In this section we derive an analytical expression for the vector field in the close vicinity to the point singularity which allowed us to identify the role of Dzyaloshinskii-Moriya interaction in its energetics. 

It is easy to show that the model Hamiltonian Eq. \eqref{E_tot} of the main text can be reduced in continuum approximation to the following functional \cite{Bar,Bak}: 
\begin{equation}
\mathcal{E} = \int_V  
w_\mathcal{E}({\bf r})  d{\bf r} , \quad\quad w_\mathcal{E}( {\bf r}) =  \mathcal{A} \left( {{\partial}_x\mathbf{n}}^2 + {{\partial}_y\mathbf{n}}^2 + {{\partial}_z\mathbf{n}}^2   \right) +  
\mathcal{D} \left(  \mathbf{n}\cdot \mathrm{rot}\,\mathbf{n}   \right) + 
H M_\mathrm{s} (1-{n}_z) 
\label{W_e}
\end{equation} 
where $\mathbf{n} \equiv \mathbf{n}(\mathbf{r})$ is a continuous unit vector field defined everywhere except at the singular points. $\mathcal{A}$ and $\mathcal{D}$ are micromagnetic constants for exchange and Dzyaloshinskii-Moriya interactions, respectively, $H$ is the applied magnetic field, and $M_\mathrm{s}$ is the magnetization of the material (the total magnetic dipole moment per unit volume). 
For the case of a simple cubic lattice micromagnetic constants are related to the microscopic constants of the discrete model as 
\begin{equation}
\mathcal{A}=J/(2a), \mathcal{D}=D/a^2, M_\mathrm{s}=\mu_\mathrm{s}/a^3,
\end{equation} 
where $a$ is the lattice constant, $\mu_\mathrm{s}$ is the magnetic dipole moment of the magnetic atom. 

In spherical coordinate parametrisation, the components of the unit vector field are defined as 
\begin{center}
$\mathbf{n} = ( \sin(\Theta) \cos(\Phi), \sin(\Theta) \sin(\Phi), \cos(\Theta)    )$,
\end{center}
where $\Theta\equiv \Theta(\mathbf{r})$ and $\Phi\equiv \Phi(\mathbf{r})$. 

To minimize the functional Eq. \eqref{W_e} one needs to solve corresponding Euler-Lagrange equations which in this case has a following form:
\begin{equation}
\begin{cases}
\mathcal{A} \left(  \sin( 2 \Theta) (\nabla\Phi)^2 - 2 \Delta\Theta \right) 
- 2 \mathcal{D} \sin(\Theta) \left( {\mathbf{n}}\cdot \nabla\Phi  \right) 
+ H M_\mathrm{s} \sin(\Theta)  
 =  0, \\
- \mathcal{A} \left( 2 \cos(\Theta) (\nabla\Theta \cdot \nabla\Phi) + \sin(\Theta) \Delta\Phi   \right)
+ \mathcal{D}  \left( {\mathbf{n}}\cdot \nabla\Theta  \right)   
 =  0.
\end{cases}
\label{eq:EL}
\end{equation}

Let's assume that the singular point coincide with the origin of spherical coordinate system $(r,\vartheta,\varphi)$.
Then, in the close vicinity to this point, the Taylor expansion of $\Theta(\mathbf{r})$ and $\Phi(\mathbf{r})$ reads
\begin{align}
\Theta &= \sum\limits_{i=0}^\infty t_i(\vartheta,\varphi)\left( \frac{\mathcal{D}}{\mathcal{A}} \,  r  \right)^i, \label{eq:ser1} \\ 
\Phi &= \sum\limits_{i=0}^\infty f_i(\vartheta,\varphi)\left( \frac{\mathcal{D}}{\mathcal{A}} \,  r  \right)^i. \label{eq:ser2}
\end{align}
In the assumption that magnetization distribution around the singular point has an axially symmetric structure, the first three terms in Eqs.(\ref{eq:ser1}, \ref{eq:ser2}), which satisfy the Eqs. (\ref{eq:EL}) are
\begin{align}
t_0 &= 2 \arctan \left(c \cot(\vartheta/2) \right), \nonumber \\
t_1 &= - \frac{\sin( \varphi_1)}{2} \sin( \vartheta ),  \nonumber\\
t_2 &= \frac{ (1-c^2)\cos(\varphi_1)^2}{16 c}\sin( \vartheta ) -
\frac{ \left(2 c + (1+c^2)\cos(\varphi_1) \right)^2}{32 c (1+c^2)}\sin( 2\vartheta ) + \nonumber \\
& \quad + \left( \frac{c (1-c^2)}{12 (1+c^2)} + \frac{c \mathcal{A} H M_\mathrm{s}}{6 \mathcal{D}^2} \right) \frac{ \cot(\vartheta/2) }{1 + {c}^2 \cot(\vartheta/2)^2 }, \label{eq:ser_t} 
\end{align}
and
\begin{align}
f_0 &= \varphi + \varphi_1, \nonumber \\
f_1 &= - \frac{(1-c^2)\cos( \varphi_1)}{4{c}} + \frac{2 c+(1+c^2)\cos( \varphi_1)}{4 c} \cos( \vartheta ),  \nonumber \\
f_2 &= -\frac{\sin \left(\varphi_1\right) \left(9 \left(c^6+c^4+c^2+1\right) \cos \left(\varphi_1\right)+6 c^5-8 c^3+6 c\right)}{96 c^2 \left(c^2+1\right)} + \nonumber \\
& \quad +\frac{ \left(1 - c^2\right) \sin \left(\varphi_1\right) \left(\left(c^2+1\right) \cos \left(\varphi_1\right)+c\right)}{8 c^2} \cos(\vartheta) + \nonumber \\ 
& \quad + \frac{\left(-1-c^4\right) \sin \left(\varphi_1\right) \left(\left(c^2+1\right) \cos \left(\varphi_1\right)+2 c\right)}{32 c^2 \left(c^2+1\right)} \cos(2 \vartheta), \label{eq:ser_f}
\end{align}
where $c$ and $\varphi_1$ are arbitrary free constants, which are not defined by the asymptotic expansion but they can be determined numerically.
Note, the assumption of an axisymmetric structure, strictly speaking, is satisfied only at magnetic field above the saturation of the conical phase, $H \geq H_D\equiv{\mathcal{D}^2}/({2M_\textrm{s}\mathcal{A}})$.

Using the Eqs. (\ref{eq:ser1} - \ref{eq:ser_f}) we have got an expression for the energy contribution from the volume inside the sphere of radius $R$ surrounding the singularity:
\begin{equation}
\mathcal{E}_\mathrm{s}(R) = \int\limits_{0}^{R}\int\limits_{0}^{\pi}\int\limits_{0}^{2\pi} w_\mathcal{E}({\bf r})\, r^2\!\sin(\vartheta)   d\varphi d\vartheta dr = 8 \pi \mathcal{A} R + \frac{2 \pi \mathcal{D}^2}{ 3 \mathcal{A} } \beta R^3 +\mathcal{O}(R^4),
\label{E_s}
\end{equation}
where
\begin{equation}
\beta = -\frac{ (1 + {c}^2 + 2 c \cdot \cos(\varphi_1)) (1-{c}^4 + 4 {c}^2 \ln(c) )}{ (1 - {c}^2)^3} - \frac{4 \mathcal{A} H M_\textrm{s} {c}^2  (1 - {c}^2 + 2 \ln(c))} {\mathcal{D}^2 (1 - {c}^2)^2}.
\nonumber
\end{equation}
In particular, for $H = H_D$, the domain of definition for $\beta$  is $-1.06 < \beta < 1.32$, but the exact value can be derived only by numerical methods.
Here it is important to note, that the constants $c$ and $\varphi_1$, in fact, are strictly dependent on the ratio $H/H_\mathrm{D}$. 

In the limiting case $H=0$ and $\mathcal{D}\rightarrow 0$, the expression in Eq.\eqref{E_s} shows a linear dependence on $R$ and converges to the energy dependence derived earlier for singularity in pure ferromagnets \cite{Doring}.
Thus, the energy contribution to the total energy related to the small volume around  the singularity resulting from the Dzyaloshinskii-Moriya interaction is proportional to $R^3$.
From above one may conclude that the energetics of singular points in chiral magnets is more intricate than in pure ferromagnets.

\section{Direct energy minimization technique}%------------------------------

For the most general case discussed in the main text the discrete model Hamiltonian is
\begin{align}
E\!= E_0  - J \sum_{\left\langle ij\right\rangle }
 \mathbf{n}_i  \cdot  \mathbf{n}_j - 
\sum_{\left\langle ij\right\rangle } 
\!\mathbf{D}_{ij}  \cdot  [\mathbf{n}_i\! \times\!  \mathbf{n}_j]   - \! \mu_\mathrm{s} \mathbf{H}\!\sum_{i}\mathbf{n}_i
-\!  K_\mathrm{u} \sum_{i}
 {n}^2_{i,\mathrm{z}}
-  K_\mathrm{c} \sum_{i}  \left( n^4_{i,\mathrm{x}} + n^4_{i,\mathrm{y}} + n^4_{i,\mathrm{z}} \right) 
\label{eq:E_tot}
\end{align}
with the assumed constraint 
\begin{equation}
n^2_{i,\mathrm{x}} + n^2_{i,\mathrm{y}} + n^2_{i,\mathrm{z}} = 1,
\label{eq:con}
\end{equation}
which means that the order parameter space is a two-dimensional sphere $\mathbb{S}^2$ and each vector $\mathbf{n}_i$ is considered as a point on it, see Fig.~\ref{Stereo}. 
For the case of such constraint a few different minimization technique have been proposed \cite{Fadd1997, Goldfarb, BatSut}. Here we use nonlinear conjugate gradient (NCG) method with  Polak-Ribi\`{e}re formula for updating  of the conjugate direction.
To conserve the constrain (\ref{eq:con}) it is convenient to use, 
instead of $n_{\mathrm{x}}$, $n_{\mathrm{y}}$,$n_{\mathrm{z}}$, the  stereographic projection defined by the pair of corresponding coordinates $\gamma_1$, $\gamma_2$, see Fig.~\ref{Stereo} .
Such a stereographic projection is defined for each point on the sphere, except at the sphere's poles from which the points are projected onto the target plane.
The projection of the point with $\theta\rightarrow0$ ($\theta\rightarrow\pi$) is not defined for the projection from the north pole (from the south pole) because $\sqrt{\gamma^2_1+\gamma^2_2}\rightarrow \infty$.
This is quite inconvenient for a computer implementation where the  values of the variables are always restricted to a finite range.
Here we propose an alternative approach based on \textit{adaptive stereographic projections} which allows to overcome the problem of the poles inertia \cite{Fadd1997}.
Contrary to one pole stereographic projection, we consider both type of projections: from the north pole
\begin{equation}
n_{\mathrm{x}} = \frac{2 \gamma^{(N)}_1}{1+(\gamma^{(N)}_1)^2+(\gamma^{(N)}_2)^2}, \quad 
n_{\mathrm{y}} = \frac{2 \gamma^{(N)}_2}{1+(\gamma^{(N)}_1)^2+(\gamma^{(N)}_2)^2}, \quad 
n_{\mathrm{z}} = \frac{-1+(\gamma^{(N)}_1)^2+(\gamma^{(N)}_2)^2}{1+(\gamma^{(N)}_1)^2+(\gamma^{(N)}_2)^2}
\label{eq:N_pro}
\end{equation}
and from the south pole:
\begin{equation}
n_{\mathrm{x}} = \frac{2 \gamma^{(S)}_1}{1+(\gamma^{(S)}_1)^2+(\gamma^{(S)}_2)^2}, \quad 
n_{\mathrm{y}} = \frac{2 \gamma^{(S)}_2}{1+(\gamma^{(S)}_1)^2+(\gamma^{(S)}_2)^2}, \quad 
n_{\mathrm{z}} = \frac{1-(\gamma^{(S)}_1)^2-(\gamma^{(S)}_2)^2}{1+(\gamma^{(S)}_1)^2+(\gamma^{(S)}_2)^2}
\label{eq:S_pro}
\end{equation}
\begin{Figure}
\centering
\includegraphics[width=0.5\columnwidth]{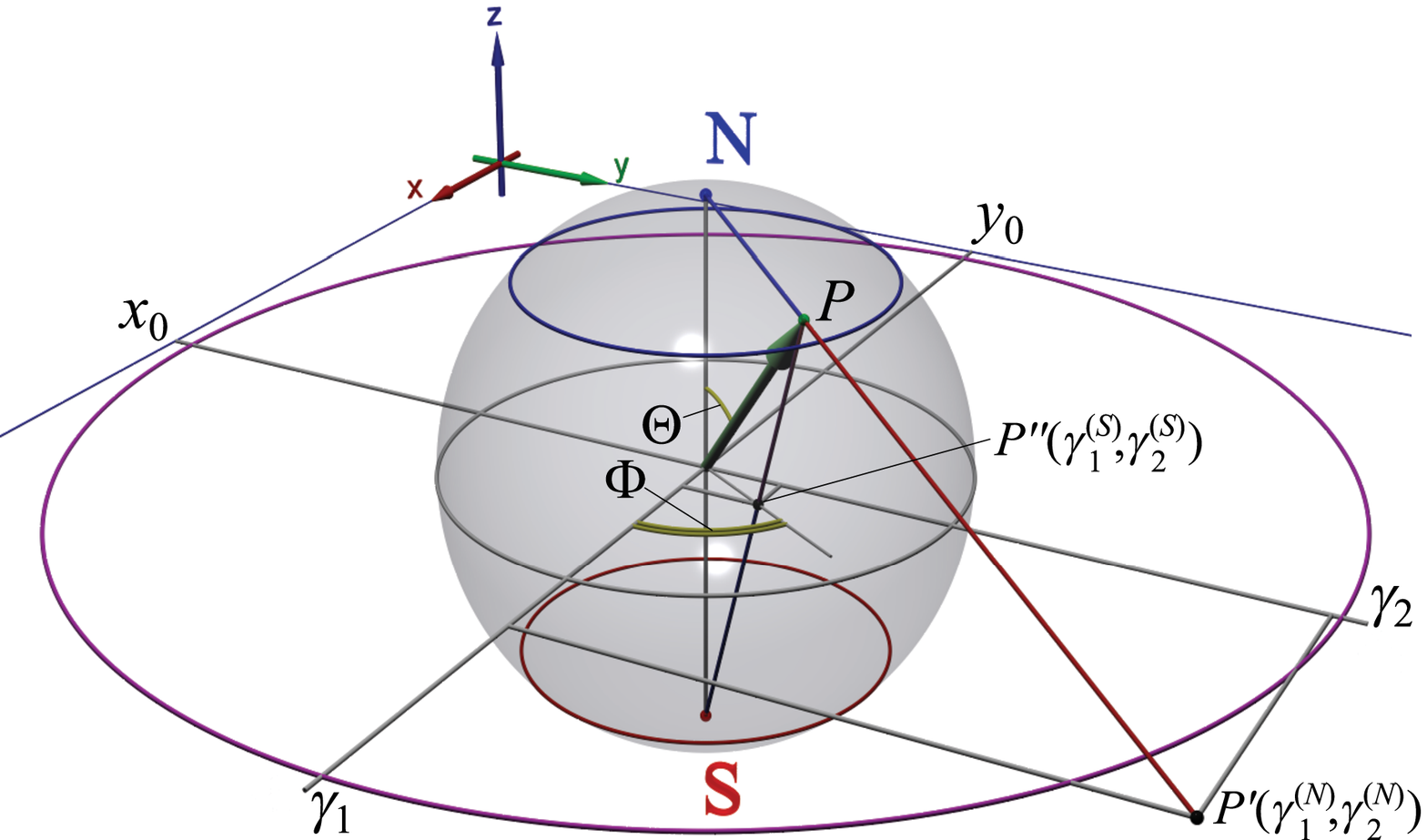}
\caption{\small Stereographic projection from the unit sphere onto the equatorial plane $z = 0$. Point $P$ on the sphere can be projected from the north pole $N$  to point $P^{\prime}$ or alternatively from the south pole $S$ to point $P^{\prime\prime}$ onto the target plane.
}
\label{Stereo}
\end{Figure}

Corresponding coordinate homeomorphisms (inverse formulas) read:
\begin{equation}
\mathbb{S}^2 \setminus \{N\} \rightarrow \mathbb{R}^2 : 
\begin{cases}
\gamma^{(N)}_1 = n_{\mathrm{x}}/(1-n_{\mathrm{z}}) \\
\gamma^{(N)}_2 = n_{\mathrm{y}}/(1-n_{\mathrm{z}})
\label{eq:N_hom}
\end{cases}
\end{equation}
\begin{equation}
\mathbb{S}^2 \setminus \{S\} \rightarrow \mathbb{R}^2 : 
\begin{cases}
\gamma^{(S)}_1 = n_{\mathrm{x}}/(1+n_{\mathrm{z}}) \\
\gamma^{(S)}_2 = n_{\mathrm{y}}/(1+n_{\mathrm{z}})
\label{eq:S_hom}
\end{cases}
\end{equation}

From the topology point of view, such a set of two coordinate charts is an atlas for $\mathbb{S}^2$.
In our implementation of the NCG method, we allow an independent switching between the two types of projections for each spin, see Supplementary movie 4 \cite{Supplement}. 
In particular, for initial configuration we use north pole projection Eqs.~(\ref{eq:N_pro}) for points with initial value $n_z<0$ and south pole projection Eqs.~(\ref{eq:S_pro}) for those which have $n_z\geq0$, see Fig.~\ref{spheres}a. 
On each NCG iteration step each spin accepts new coordinates with a small increment with respect to the previous state. 
Some spins which are initially situated in the southern semisphere can enter the northern semisphere and vice versa, see Fig.~\ref{spheres}b. 
At some iteration one or few points may appear too far away from its initial semisphere, see Fig.~\ref{spheres}c. 
The criterion for that can be \textit{e.g.} $\sqrt{\gamma^2_1+\gamma^2_2}>\Gamma_\mathrm{c}$, where $\Gamma_\mathrm{c}$ is some arbitrary chosen value. 
In our implementation we choose it such that $\sqrt{\gamma^2_1+\gamma^2_2}=\Gamma_\mathrm{c}$ which corresponds to $n_z=0.6$ for the spins defined by the north pole and $n_z=-0.6$ for south pole projection respectively.   
As soon as such event occurs at least for one spin, then all the points are switched to other set of variables in the same way as at initial step, see Fig.~\ref{spheres}d. Thus some coordinates $(\gamma^{(N)}_1,\gamma^{(N)}_2)$ are changed into corresponding $(\gamma^{(S)}_1,\gamma^{(S)}_2)$ and vice versa:
\begin{equation}
\gamma^{(N)}_1 = \frac{\gamma^{(S)}_1} {(\gamma^{(S)}_1)^2 + (\gamma^{(S)}_2)^2}, \quad \gamma^{(N)}_2 = \frac{\gamma^{(S)}_2} {(\gamma^{(S)}_1)^2 + (\gamma^{(S)}_2)^2},
\label{eq:switchN}
\end{equation}  
\begin{equation}
\gamma^{(S)}_1 = \frac{\gamma^{(N)}_1} {(\gamma^{(N)}_1)^2 + (\gamma^{(N)}_2)^2}, \quad \gamma^{(S)}_2 = \frac{\gamma^{(N)}_2} {(\gamma^{(N)}_1)^2 + (\gamma^{(N)}_2)^2}.
\label{eq:switchS}
\end{equation} 
Then the NCG starts again with the new set of stereographic coordinates $\gamma_1$, $\gamma_2$.
Note that during the minimization process such a global switching occures very seldom or never, it always depend on the initial configuration.

\begin{Figure}
\centering
\includegraphics[width=0.8\columnwidth]{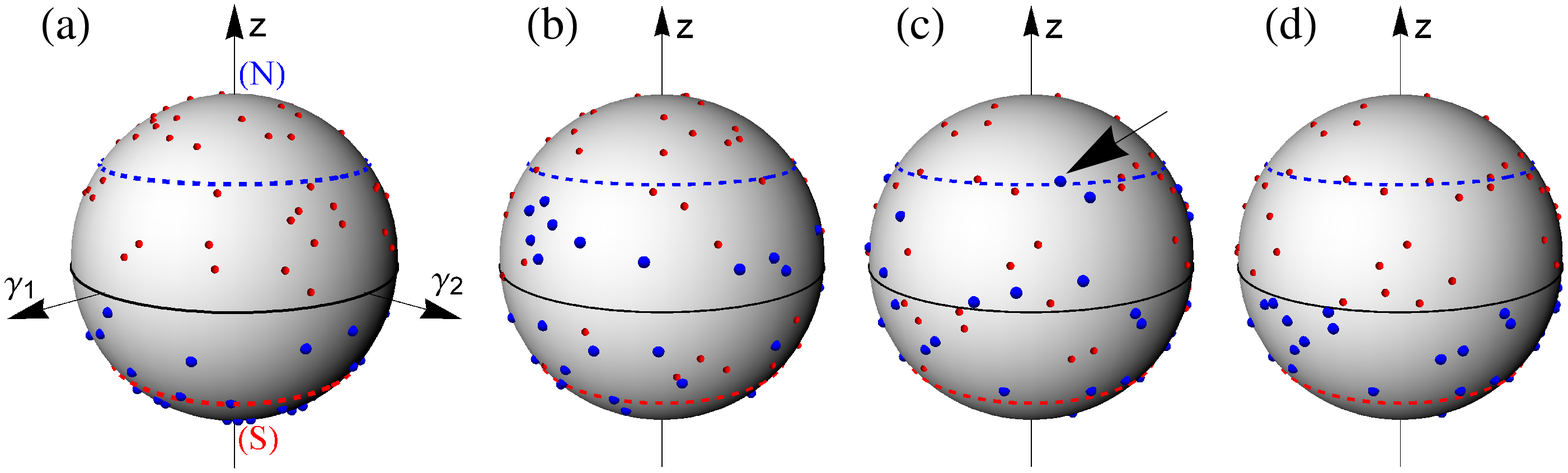}
\caption{\small Illustration for the adaptive algorithm of the switching between stereographic projections. Big blue dots are spins defined by stereographic projection from the north pole, small red dots are spins defined by projection from the south pole, see also Fig.~\ref{Stereo}. Dotted curves define critical distances from the pole of projection, see definition for $\Gamma_\mathrm{c}$ in the text. (a) the initial state, (b) the state after several iterations using the NCG method, (c) the state at which one marked point has exceeded the critical distance, (d) the same state as (c) but with redefined variables of the stereographic projections, see Eq.~(\ref{eq:switchN}, \ref{eq:switchS}).}
\label{spheres}
\end{Figure}

In order to distinguish $(\gamma^{(N)}_1,\gamma^{(N)}_2)$ from $(\gamma^{(S)}_1,\gamma^{(S)}_2)$ without introducing additional logical variable which define the type of corresponding projection for each spin e.g. $N$ (true) and $S$ (false), we used the following trick.
In our stereographic projection scheme the center of the unit sphere is assumed to be located at some arbitrary position $(x_0$, $y_0)$ such that $\sqrt{x^2_0+y^2_0}>\Gamma_\mathrm{c}$ and $x_0>0$, $y_0>0$.
Thereby, the sign of the variables $\gamma_1$ ,$\gamma_2$ stored in the computer memory can be used as the indicator for the type of the projection,
while in Eqs.(\ref{eq:switchN},\ref{eq:switchS}) the absolute values of the variables are used and a shift of the basis (see $x_0,y_0$ in Fig.\ref{Stereo}) has to be taken into account. 

Such an adaptive stereographic projection scheme i) allows one to obtain supreme performance of the NCG method, ii) fully satisfies the constraint Eq.~ (\ref{eq:con}), and iii) allows to avoid the problem with the inertia of the poles \cite{Fadd1997}. 
Contrary to other methods, see e.g. Refs.~\cite{Goldfarb, BatSut}, another important advantage of our approach is the absence of any free scalar parameter which would require an additional strategy to replace this free parameter by the optimal value to archive the best performance.

The algorithm has been implemented in a NVIDIA CUDA architecture for massive parallel calculations and real time visualization and has been used for the calculations on the graphic cards with microprocessors GK104 and GK110B. 
In order to achieve maximum performance we used single-precision floating-point format and implemented Kahan summation and optimized parallel reduction algorithm \cite{Harris} in order to reduce numerical errors.

As initial conditions we used cone phase with vortex-like cylindrical inset:
\begin{equation}
\begin{cases}
\Theta_\mathrm{ini} = \pi \left(1 - \sqrt{x^2+y^2}/{L_D} \right), \\
\Phi_\mathrm{ini} = \arctan({y}/{x} ) + {\pi}/{2} 
\end{cases}
\quad (\sqrt{x^2+y^2}<L_D)
\end{equation}
where $L_\mathrm{D}=2\pi a J/|D_{ij}|$, see also Eq.\eqref{LH_params} in the main text.
For the calculation of the chiral bobber with the NCG method, the height of the cylinder has been restricted by $0.65 L_\mathrm{D}$, while the thickness $L$ of the simulated magnetic film has been varied in the range between $1L_\mathrm{D}$ and $7L_\mathrm{D}$.

\section{Monte-Carlo simulations}%-------------------------------------
In order to identify the probability for spontaneous nucleation of the chiral bobber driven by ordinary thermal fluctuations we also perform simulated annealing \cite{Laarhoven} with a Monte Carlo method.
We implemented the classical Metropolis algorithm \cite{MCS}. 
In the simulation of the annealing process we use an exponential rule for the temperature sequence:
\begin{equation}
T_{i+1} = 0.95 \, T_{i}.
\end{equation}
In the Supplementary Movie 3 \cite{Supplement}  we present real-time simulations of the annealing process, performed for the following set of parameters: $J=1$, $D=0.15$, $\mu_\mathrm{s} H=0.018$ and domain size of $128 \times 256 \times 100$ spins. Detailed description of the movie is presented below.

The simulation starts with a random spin distribution and an initial temperature of $T_0= 3 J/ k_\mathrm{B}$, above the Curie temperature for the used set of system parameters.   
The temperature was gradually reduced down to $T=0.05 J/ k_\mathrm{B}$ and then the NCG minimizer has been used instead of the Monte Carlo iterations in order to complete the full relaxation process faster. 
It is done only for the reason of better visualization.
Performing the full relaxation with Monte Carlo requires much more iterations than the relaxation with NCG minimizer.

At $T=0.6J/ k_\mathrm{B}$ on the top surface on the right hand side of the simulated domain one can already see a few bright spots corresponding to the nucleated chiral bobbers, which is illustrated by scanning through the thickness.
After the full relaxation the spin structure shows the conical state with three chiral bobbers inside, two on the top and one on the bottom surface.
Finally we show iso-surfaces with $n_z=\mathrm{const}$, which illustrate the localization and the relative positions of the chiral bobbers. 

The number of Monte Carlo steps as well as the variation in the domain size and coupling constants obviously affects the probability for the chiral bobber nucleation. 
As far as we pursuit the purpose to find the regime for metastable state nucleation we used relatively fast cooling with a reduced number of Monte Carlo steps as compared to the slow cooling regime when the purpose is to find the ground state.
With the set of parameters mentioned above we have found the optimal number of Monte Carlo steps to be 75 per spin at each temperature step which allows to observe the spontaneous nucleation of the chiral bobber with the relatively high probability.
However, even for this fast cooling regime the spontaneous nucleation of metastable skyrmion tube has not been observed in our simulations.

\section{Geodesic nudged elastic band method}\label{GNEB}%-------------------

\begin{Figure}
\centering
\includegraphics[width=0.8\columnwidth]{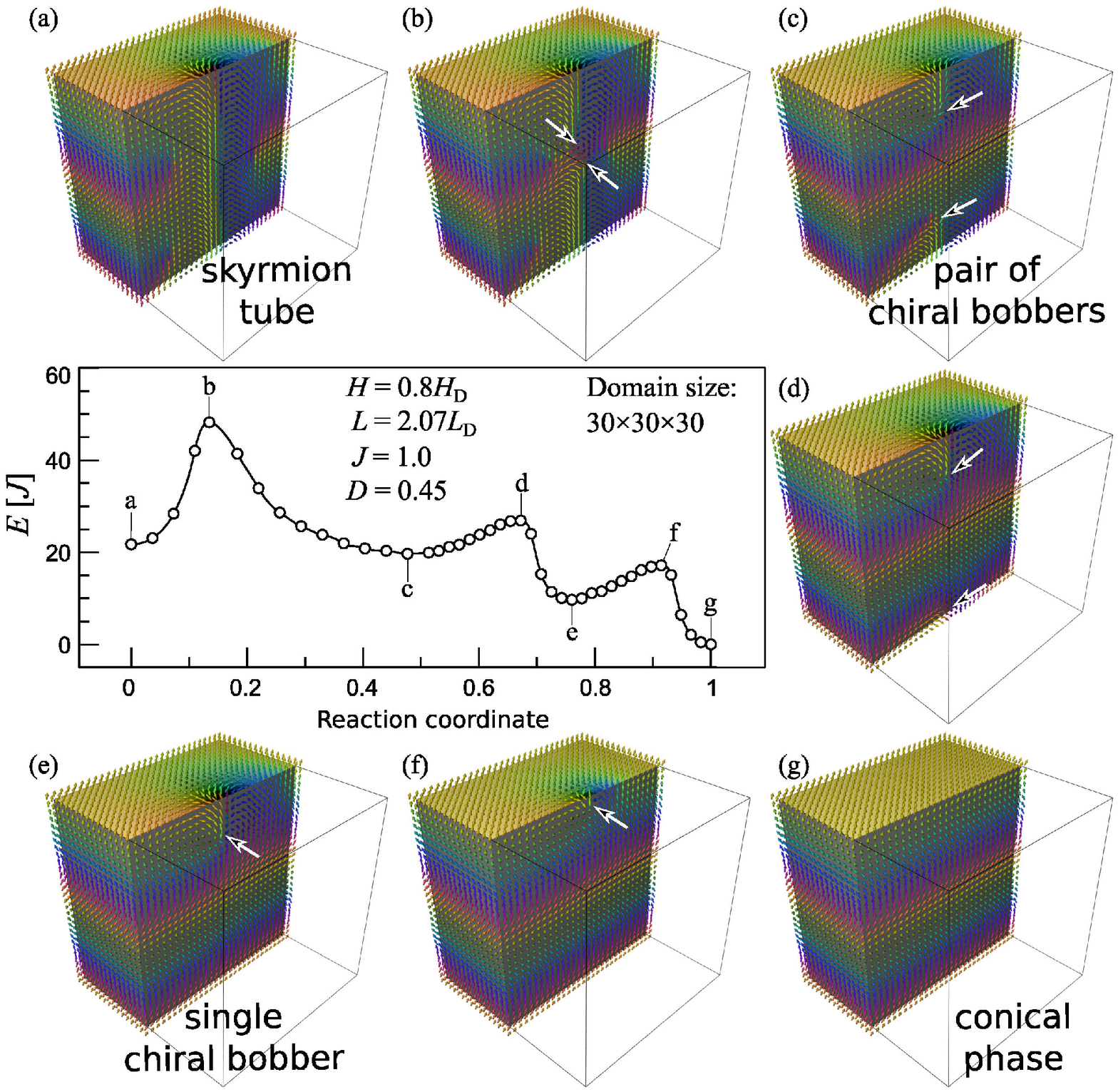}
\caption{\small The minimum energy path (MEP) identical to presented in the main text, Fig.~\ref{MEP}. 
Insets show images corresponding to the extrema on the MEP marked with $a-g$.
(a) isolated skyrmion tube, (b) nucleation of two chiral Bloch points, (c) metastable pair of chiral bobbers, (d) collapse of chiral bobber located on bottom surface via pushing out the chiral Bloch point to the bottom surface, (e) the single chiral bobber, (f) collapse of the single chiral bobber, (g) the conical ground state. Arrows indicate approximate position of the chiral Bloch point.
\label{MEP-supl}}
\end{Figure}

In the calculation of the minimum energy path (MEP) with the geodesic nudged elastic band (GNEB) method we follow exactly the algorithm in Ref. \cite{Bessarab_15sup}. 
We started with the relaxed isolated skyrmion and the pure conical state
which are used as initial and final configurations, respectively, for the desired discrete MEP.
We used a linear interpolation with small stochastic noise to set intermediate states and construct an initial path between the initial and final configurations.

The typical number of intermediated states (or images) was set to 15.
Instead of the velocity projection optimization on a sphere \cite{Bessarab_15sup} we used another method \cite{Suess_07} which we found to be more convenient for our implementation, but possibly less efficient.
In particular, for each path optimization step of each spin at the lattice site $i$ of image $k$, we solve the following partial differential equation 
\begin{equation}\label{gneb:1}
\frac{\partial \mathbf{n}_{k,i}}{\partial t}=
-\alpha \mathbf{n}_{k,i} \times [\mathbf{n}_{k,i} \times \mathbf{D}_{k,i}],
\end{equation}
which is equivalent to the Landau-Lifshitz-Gilbert equation containing only the damping term, but with the recession term totally neglected.
The $3N$-dimensional vector $\mathbf{D}_{k}=(\mathbf{D}_{k,1},\mathbf{D}_{k,2},...,\mathbf{D}_{k,N})$ ($N$ is the number of spins) in Eq. \eqref{gneb:1} controls the relaxation of each image towards the MEP and is defined as
%\refstepcounter{equation}
\begin{equation}\label{gneb:2}
\mathbf{D}_{k}=\left(\mathbf{H}_{\mathrm{eff},k}(\mathbf{n}_{k})
-(\mathbf{H}_{\mathrm{eff},k}(\mathbf{n}_{k})\cdot \hat{\mathbf{\tau}}_{k})\hat{\mathbf{\tau}}_{k}  \right) + \mathbf{F}_{k},
%\tag{S\Roman{section}.\arabic{equation}}
\end{equation}
where $\mathbf{H}_{\mathrm{eff},k}$, $ \mathbf{F}_{k}$, and $\hat{\mathbf{\tau}}_{k}$ are $3N$-dimensional vectors of the effective field, the so-called spring force, and the normalized tangent vector laying in the tangent space of image $k$, respectively. For detail see Ref.\cite{Bessarab_15sup}.
Following the algorithm Ref.\cite{Bessarab_15sup} we used path splitting between nearest metastable states and relaxed each of local minima independently.
In Fig.\ref{MEP-supl} we present the same MEP as in the main text together with exact view of the images corresponding to extrema on the MEP.

\renewcommand{\section}[2]{}


\begin{thebibliography}{99}
\bibitem{Parkin_15}
S. Parkin, S-H. Yang,  %Memory on the racetrack. 
Nature Nanotech. \textbf{10}, 195 (2015).%doi:10.1038/nnano.2015.41

\bibitem{Bogdanov_89}
A. N. Bogdanov, D. A. Yablonskii,  
%Thermodynamically stable "vortices" in magnetically ordered crystals. The mixed state of magnets. 
Sov. Phys. JETP \textbf{68}, 101 (1989).

\bibitem{Kiselev_11}
N. S. Kiselev, A. N. Bogdanov, R. Sch\"{a}fer,  U. K. R\"{o}{\ss}ler, 
%Chiral skyrmions in thin magnetic films: new objects for magnetic storage technologies? 
J. Phys. D: Appl. Phys. \textbf{44}, 392001 (2011).

\bibitem{Fert_13}
A. Fert, V. Cros, J. Sampaio,  
%Skyrmions on the track. 
Nature Nanotech. \textbf{8}, 152 (2013).

\bibitem{Dzyaloshinskii}
I. Dzyaloshinskii, Journal of Physics and Chemistry of Solids \textbf{4}, 241 (1958).

\bibitem{Moriya}
T. Moriya,  
%Anisotropic superexchange interaction and weak ferromagnetism. 
Phys. Rev. \textbf{120}, 91 (1960).

\bibitem{Rajaraman}
R. Rajaraman, \textit{Solitons And Instantons: An Introduction To Solitons And Instantons In Quantum Field Theory} (North-Holland, 1982).

\bibitem{Bogdanov_95}
A. Bogdanov, JETP Lett. \textbf{62}, 247 (1995).
% [A. Bogdanov, Pisma v Zh. Eksp. Teor. Fiz. \textbf{62}, 231 (1995)].

\bibitem{Shibata_13}
K. Shibata, X. Z. Yu, T. Hara, D. Morikawa, N. Kanazawa, K. Kimoto, S. Ishiwata, Y. Matsui, Y. Tokura,
%Towards control of the size and helicity of skyrmions in helimagnetic alloys by spin–orbit coupling
Nature Nanotech. \textbf{8}, 723 (2013).

\bibitem{Romming_13}
N. Romming, M. Menzel, C. Hanneken, J. E. Bickel, B. Wolter, K. von Bergmann,  A. Kubetzka, R. Wiesendanger,  
%Writing and deleting single magnetic skyrmions. 
Science \textbf{341}, 636 (2013).

\bibitem{Yu_10}
X. Z. Yu, Y. Onose,  N. Kanazawa, J. H. Park,  J. H. Han, Y. Matsui,  N. Nagaosa, Y. Tokura,  
%Real space observation of a two-dimensional skyrmion crystal.
Nature \textbf{465}, 901 (2010).

\bibitem{Bogdanov_94}
A. Bogdanov, A. Hubert,  
%Thermodynamically stable magnetic vortex states in magnetic crystals. 
J. Magn. Magn. Mater. \textbf{138}, 255 (1994).

\bibitem{Bogdanov_99}
A. Bogdanov, A. Hubert, 
% The stability of vortex-like structures in uniaxial ferromagnets. 
J. Magn. Magn. Mater. \textbf{195}, 182 (1999).

\bibitem{Bogdanov_06}
U. K. R\"{o}{\ss}ler, A. N. Bogdanov,   C. Pfleiderer, %Spontaneous skyrmion ground states in magnetic metals. 
Nature \textbf{442}, 797 (2006).

\bibitem{Bogdanov_11}
U. K. R\"{o}{\ss}ler, A. A. Leonov, A. N. Bogdanov,  
%Chiral Skyrmionic matter in non-centrosymmetric magnets. 
J. Phys.: Conf. Ser. \textbf{303}, 012105 (2011).

\bibitem{Supplement}
See Appendices for details. See \href{http://www.youtube.com/channel/UCN4mgZGR4Yv3T9Yc-94RDAA}{\textbf{\underline{YouTube channel}}} for exact spin structure corresponding to the skyrmion tube (Movie 1) and chiral bobber (Movie 2), visualization of Monte Carlo simulation for thermally activated nucleation of chiral bobber (Movie 3). {\tiny http://www.youtube.com/channel/UCN4mgZGR4Yv3T9Yc-94RDAA}



\bibitem{helix}
I. E. Dzyaloshinskii, 
%Theory of helicoidal structures in antiferromagnets. III,
Sov. Phys. JETP \textbf{20}, 665 (1965).

\bibitem{Rybakov_13}
F. N. Rybakov, A. B. Borisov,  A. N. Bogdanov, 
%Three-dimensional skyrmion states in thin films of cubic helimagnets.
Phys. Rev. B \textbf{87}, 094424 (2013).

\bibitem{Yokouchi_14}
T. Yokouchi, N. Kanazawa, A. Tsukazaki, Y. Kozuka, M. Kawasaki, M. Ichikawa, F. Kagawa, and Y. Tokura, 
Phys. Rev. B \textbf{89}, 064416 (2014).

\bibitem{Yu_15}
X. Yu, A Kikkawa, D. Morikawa, K Shibata, Y. Tokunaga, Y. Taguchi, Y. Tokura,
%Variation of skyrmion forms and their stability in MnSi thin plates
Phys. Rev. B \textbf{91}, 054411 (2015).

\bibitem{Yu_11}
X. Z. Yu, N. Kanazawa, Y. Onose, K. Kimoto, W. Z. Zhang, S. Ishiwata, Y. Matsui, Y. Tokura,  
%Near room-temperature formation of a skyrmion crystal in thin-films of the helimagnet FeGe. 
Nature Mater. \textbf{10}, 106 (2011).

\bibitem{Dupe_14}
B. Dupe, M. Hoffmann, C. Paillard, S. Heinze, 
%Tailoring magnetic skyrmions in ultra-thin transition metal films
Nature Com. \textbf{5}, 4030 (2014).

\bibitem{Bogdanov_14}
M. N. Wilson, A. B. Butenko, A. N. Bogdanov, T. L. Monchesky, 
%Chiral skyrmions in cubic helimagnet films: The role of uniaxial anisotropy. 
Phys. Rev. B \textbf{89}, 094411 (2014).

\bibitem{Leonov_PhD}
A. A. Leonov, Ph.D. thesis, 
%Twisted, localized, and modulated states described in the phenomenological theory of chiral and nanoscale ferromagnets
Technical University Dresden (2012).

\bibitem{Surf_twist1}
M. N. Wilson, E. A. Karhu, D. P. Lake, A. S. Quigley, S. Meynell, A. N. Bogdanov, H. Fritzsche, U. K. R\"{o}\ss ler, T. L. Monchesky, 
%Discrete helicoidal states in chiral magnetic thin films
Phys. Rev. B \textbf{88}, 214420 (2013).

\bibitem{Surf_twist2}
S. A. Meynell, M. N. Wilson, H. Fritzsche, A. N. Bogdanov, T. L. Monchesky, Phys. Rev. B \textbf{90}, 014406 (2014).

\bibitem{Malozemoff}
A. P. Malozemoff, J. C. Slonczewski,
\textit{Magnetic domain walls in bubble materials}
(Academic Press, New York, 1979).

\bibitem{Ray_14} M. W. Ray, E. Ruokokoski, S. Kandel, M. M\"{o}tt\"{o}nen, D. S. Hall, Nature \textbf{505}, 657 (2014).

\bibitem{Castelnovo_08} C. Castelnovo, R. Moessner, S. L. Sondhi, Nature \textbf{451}, 42 (2008).

\bibitem{Milde}
P. Milde, D. K\"{o}hler, J. Seidel, L. Eng, A. Bauer, A. Chacon, J. Kindervater, S. M\"{u}hlbauer, C. Pfleiderer, S. Buhrandt, C. Sch\"{u}tte, A. Rosch,
%Unwinding of a Skyrmion Lattice by Magnetic Monopoles
Science \textbf{340}, 1076 (2013).

\bibitem{Rosch}
C. Sch\"{u}tte, A. Rosch, 
%Dynamics and energetics of emergent magnetic monopoles in chiral magnets
Phys. Rev. B \textbf{90}, 174432 (2014)

\bibitem{Vol1978}
G. E. Volovik, Sov. Phys. JETP Lett. \textbf{28}, 59 (1979).

\bibitem{Monastyrsky} 
M. Monastyrsky, \textit{Topology of Gauge Fields and Condensed Matter} (Springer, New York, 1993).

\bibitem{Volovik} G. E. Volovik, \textit{The Universe in a Helium Droplet} (Oxford University Press, New York, 2003).

\bibitem{Bessarab_15}
P. F. Bessarab, V. M. Uzdin,  H. J\'{o}nsson, Comp. Phys. Commun., arXiv:1502.05065.
%Method for finding mechanism and activation energy of magnetic transitions, applied to skyrmion and antivortex annihilation
%http://arxiv.org/abs/1502.05065


\bibitem{jonsson_98} 
H. J\'{o}nsson, G. Mills, K. W. Jacobsen,
`Nudged Elastic Band Method for Finding Minimum Energy Paths of Transitions', in
{\it `Classical and Quantum Dynamics in Condensed Phase Simulations'},
ed. B. J. Berne, G. Ciccotti and D. F. Coker (World Scientific, 1998), page 385.

\bibitem{Henkelman_00}
G. Henkelman, H. J\'{o}nsson
%Improved tangent estimate in the nudged elastic band method for finding minimum energy paths and saddle points
J. Chem. Phys. \textbf{113}, 9978 (2000).
%http://dx.doi.org/10.1063/1.1323224

\bibitem{Park_14}
H. S. Park, X. Yu, S. Aizawa, T. Tanigaki, T. Akashi, Y. Takahashi, T. Matsuda, N. Kanazawa, Y. Onose, D. Shindo,
A.Tonomura, Y. Tokura,
%Observation of the magnetic flux and three-dimensional structure of skyrmion lattices by electron holography
Nature Nanotech. \textbf{9}, 337 (2014).
%DOI: 10.1038/NNANO.2014.52

\bibitem{Midgley}
P. A. Midgley, R. E. Dunin-Borkowski,
%Electron tomography and holography in materials science
Nature Mater. \textbf{8}, 271 (2009). 
%doi:10.1038/nmat2406
\end{thebibliography}

\begin{thebibliography}{99}

\makeatletter
\addtocounter{\@listctr}{39}
\makeatother

\bibitem{Bar}
V. G. Bar’yakhtar, E. P. Stefanovsky, Fiz. Tverd. Tela \textbf{11}, 1946 (1969) [Sov. Phys. Solid State \textbf{11}, 1566 (1970)].

\bibitem{Bak}
P. Bak, M. H. Jensen, J. Phys. C: Solid State Phys. \textbf{13}, L881 (1980).

\bibitem{Doring}
W. D\"{o}ring, J. of Appl. Phys. \textbf{39}, 1006 (1968).

\bibitem{Fadd1997}
L. Faddeev and A. J. Niemi, Nature \textbf{387}, 58 (1997).

\bibitem{Goldfarb}
D. Goldfarb, Z. Wen, and W. Yin, SIAM J. Imaging Sci. \textbf{2}, 84 (2009).

\bibitem{BatSut}
R. A. Battye, P. M. Sutcliffe, Rev. in Math. Phys. \textbf{14}, 29 (2002).

\bibitem{Harris}
M. Harris, \textit{Optimizing parallel reduction in CUDA}, NVIDIA Documentation (2007).

\bibitem{MCS}
D. P. Landau and K. Binder, {\it A Guide to Monte Carlo Methods in Statistical Physics} (Cambridge U. Press, Cambridge, 2000).

\bibitem{Laarhoven}
P.J.M. van Laarhoven, and E. H. L. Aarts, \textit{Simulated Annealing: Theory and Applications} (Dordrecht, Holland: D. Reidel, 1987). 


\bibitem{Bessarab_15sup}
P. F. Bessarab, V. M. Uzdin,  H. J\'{o}nsson, Comp. Phys. Commun., arXiv:1502.05065.
%Method for finding mechanism and activation energy of magnetic transitions, applied to skyrmion and antivortex annihilation
%http://arxiv.org/abs/1502.05065

\bibitem{Suess_07}
D. Suess, S. Eder, J. Lee, R. Dittrich, J. Fidler, J. W. Harrell, T. Schrefl, G. Hrkac, M. Schabes, N. Supper, A. Berger, Phys. Rev. B \textbf{75}, 174430 (2007).


\end{thebibliography}
\end{document}